\documentclass[oldversion]{aa}
\usepackage{epsfig}
\usepackage{graphicx}
\usepackage{epstopdf}

\begin{document}

% - new commands
\newcommand{\hi}{\ion{H}{i}~}
\newcommand{\hii}{\ion{H}{ii}~}

   \title{Evolution of spiral galaxies in modified gravity}

   \author{O. Tiret         %\inst{1}           
\and          F. Combes        %  \inst{1}  
 }

%  \offprints{O. Tiret }

   \institute{  Observatoire de Paris, LERMA, 
            61 Av. de l'Observatoire, F-75014, Paris, France   }

   \date{Received 26/09/2006/ 08/12/2006}

   \titlerunning{Evolution of spiral galaxies in modified gravity}

   \authorrunning{Tiret \& Combes}

\abstract{We compare N-body simulations of isolated galaxies performed in both frameworks of modified Newtonian dynamics (MOND) and Newtonian gravity with dark matter (DM). We have developed a multigrid code able to  efficiently solve the modified Poisson equation derived from the Lagrangian formalism AQUAL. We take particular care of the boundary conditions that are a crucial point in MOND. The 3-dimensional dynamics of initially identical stellar discs is studied in both models. In Newtonian gravity the live DM halo is chosen to fit the rotation curve of the MOND galaxy. For the same value of the Toomre parameter ($Q_T$), galactic discs in MOND develop a bar instability sooner than in the DM model. In a second phase the MOND bars weaken while the DM bars continue to grow by exchanging angular momentum with the halo. The bar pattern speed evolves quite differently in the two models: there is no dynamical friction on the MOND bars so they keep a constant pattern speed while the DM bars slow down significantly. This affects the position of resonance like the corotation and the peanut. The peanut lobes in the DM model move radially outward while they keep the same position in MOND. Simulations of (only stellar) galaxies of different types on the Hubble sequence lead to a statistical bar frequency that is closer to observations for the MOND than the DM model.

\keywords{Galaxies: general --- Galaxies: kinematics and dynamics --- Galaxies: spiral --- Galaxies: structure --- Cosmology: dark matter}
}
\maketitle

%---------------------------------------------------------------

\section{Introduction} 
As has been emphasized in the last years, the concordance 
$\Lambda$CDM cosmological model is very successful in accounting
for large-scale structure formation (e.g., Silk 2004), but 
encounters severe problems at galactic scale: in particular
the highly peaked dark matter (DM) distribution
predicted by numerical simulations (Navarro et al. 2004) is not
compatible with most observed rotation curves of galaxies 
(de Blok 2005); the predicted angular momentum of baryons condensed 
in galaxies is much too low (Steinmetz 2003), and the number of
predicted satellites around a given giant galaxy is more than an order
of magnitude larger than what is observed (Moore et al. 1999). One solution to these problems has been searched for in the energetic feedback provided either by violent star formation (e.g., Kravtsov et al. 2004)
or by an AGN (Croton et al. 2006). However, even large variations of these
parameters have not been successful in solving the problems significantly for
all galaxy types. Another kind of solution is resorting to the 
modified Newtonian dynamics (MOND), proposed by Milgrom (1983)
 as an empirical modification of gravity, when the generated acceleration
falls below a universal value $a_0 \sim 2 \times 10^{-10}$ m s$^{-2}$.
In this model, there is no DM anymore, but the visible mass
in the inner parts of galaxies produces a much boosted gravity force in
the outer parts, with a longer range effect.  Bekenstein \& Milgrom (1984)
developed a self-consistent Lagrangian theory, where the Poisson equation
is transformed into:
\begin{equation}
\label{eq:mondaqual}
\nabla [ \mu(|\nabla\Phi|/a_0) \nabla\Phi ] = 4 \pi G \rho,
\end{equation}

\noindent where $\mu(x)$ is a function that is equal to unity at large $x$
(Newtonian regime), and tends to $x$ when $x <<$ 1 in the MOND regime.
  Far in this regime, and assuming some symmetry
(spherical, cylindrical, or plane) it can be shown that the MOND acceleration
$g_M$ satisfies the relation (Brada \& Milgrom 1995):
\begin{equation}
\label{eq:deepM2N}
g_M ^2 = a_0 g_N,
\end{equation}
\noindent where $g_N$ is the Newtonian acceleration.
This model has large success at galactic scale, in particular
explaining all rotation curves of galaxies, and naturally the Tully-Fisher relation,
as developed in the excellent review by Sanders \& McGaugh (2002).

Interest has grown in the MOND theory since the proposition by
Bekenstein (2004) of a Lorentz-covariant theory (TeVeS), able to 
replace general relativity, accounting for gravitational lensing  and 
passing elementary tests of gravity in the  solar system.
Simulations have been attempted to explore the large-scale structure
formation, with encouraging results
(Knebe \& Gibson 2004; Nusser \& Pointecouteau 2006). More recently, weak lensing observations of the bullet merging cluster 1E0657-56 (Clowe et al. 2006) claim that the spatial separation between the main baryonic component (X-ray gas) and the total mass shows direct evidence for the existence of collisionless DM. They find that any modified gravity model, considering only the baryonic mass, fails to reproduce the observations. However, Angus et al. (2006) have re-analyzed these observations in the context of modified gravity and show that the data are also compatible with the Bekenstein model of MOND, in which some collisionless dark matter exists under the form of ordinary hot neutrinos of 2 eV.

 The most stringent constraints on the choice of the interpolation function
$\mu(x)$ are expected to be obtained on a small scale however.
To  better fit the rotation curve of the Milky Way, 
the function  $\mu(x)= x/(1+x)$ has been proposed by
Famaey \& Binney (2005), in place of the
empirical initial function $\mu(x)= x/(1+x^2)^{1/2}$. 
 In addition, physical constraints and the external
field effect further reduce the choice of the interpolating function 
(Zhao \& Famaey 2006).

Since the motivation of MOND and its best success concern
the galactic scales, and in particular the rotation curves fit without
dark matter, more physical constraints should be explored at these
scales. In particular, the stability of spiral galaxies in this model,
the secular evolution taking into account spiral waves and bars have
to be investigated, to compare the dynamical behavior of a typical
galaxy in the Newtonian CDM model and the MOND frame.
Brada \& Milgrom (1999, hereafter BM99) have begun to tackle
this problem, and have shown that the Toomre Q-parameter could 
be chosen lower than in the Newtonian case, to obtain
the same stability level. The modified acceleration provided
a comparable stability level with respect to bars as does 
a dark matter halo in the Newtonian case. There are, however,
limitations in their model, since they considered infinitely thin
discs and ignored the z-structure, acceleration, and dynamics,
which are very different in Newtonian and MOND regimes.

In this work, we present numerical simulations of 
several spiral galaxy models, representing  the whole Hubble sequence
and a large mass range, in both CDM Newtonian and MOND 
models.  The goal is to find specific tests and 
constraints to the gravity theory, to be applied on a global statistical basis
and confront them to the observations.  The diagnostics are
to be found in the bar frequency, the spiral morphology, the 
thickness of discs and their box/peanut shapes, 
the surface density profiles, and the angular momentum distribution.
In this first approach, pure stellar discs are considered, while 
gas and star formation will be investigated in a future work. In the next section, we described the numerical code developed to solve the difficult problem of MOND
dynamics, and in Sect. 3 the analysis and diagnostics 
we applied to the simulations results. Initial conditions for 
spiral galaxies described in Sect. 4, are selected to be as close as possible
in the plane for the two compared models: in particular
they have the same radial baryonic distribution and the same
rotation curve and velocity dispersion. Results are
presented in Sect. 5 and then discussed in Sect. 6 to emphasize
the fundamental differences in galaxy evolution for the
two competing dynamics.

\section{Numerical model}

The non-linearity of the MOND gravity leads us to use different techniques than the usual ones for the potential solver (or force solver).

\subsection{Multigrid (MG) potential solver}

The modified Poisson equation is a non-linear elliptic partial differential equation (PDE). This kind of equation can be solved efficiently using multigrid (MG) techniques. We have written an N-body code in which we implemented a full multigrid algorithm (FMG) with full approximation scheme (FAS) for the potential solver (see Numerical Recipes, Press et al., 1992). Brada and Milgrom (BM99) used such a code to solve (\ref{eq:mondaqual}).

Up to some point, the code works like a particle-mesh(PM) code. Particles evolve in a 3D Cartesian grid. Density is computed using the cloud in cell interpolation, the potential is deduced by MG techniques, the equation of motion is solved by the leapfrog scheme. The only difference from a classical PM-code occurs in the potential solver. 
\begin{figure}
	\centering
	\includegraphics[width=6cm]{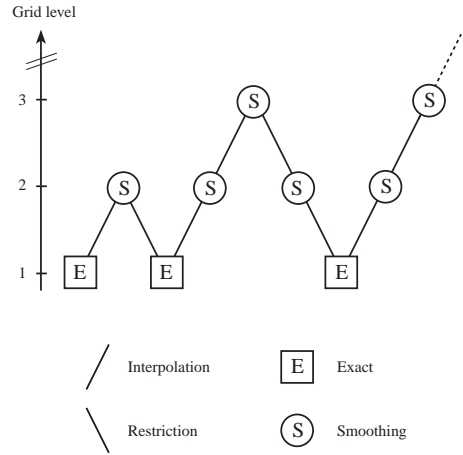}
	\caption{Full multigrid (FMG) algorithm is used to accelerate the convergence in the resolution of the modified Poisson equation (see text).}
	\label{fig:FMG}
\end{figure}

The MG computes the solution on finer and finer grids (Fig. \ref{fig:FMG}) by calculating correction terms on each level and converges even more quickly than by solving the same equation directly on the finest grid. We use the Gauss-Seidel relaxation with red and black ordering (Press et al. 1992) to solve the system of equations obtained by discretisation. This step is called \textit{smoothing}. To go from the grid level $n$ to $n+1$ we make a tri-linear interpolation (prolongation operator, P), and inversely, the full-weighting operator (R) is used to go to the level $n+1$ to $n$. The number of pre-/post-relaxations were chosen to $\nu_{pre}=2$ and $\nu_{post}=1$.

\begin{figure}
	\centering
	\includegraphics[width=5cm]{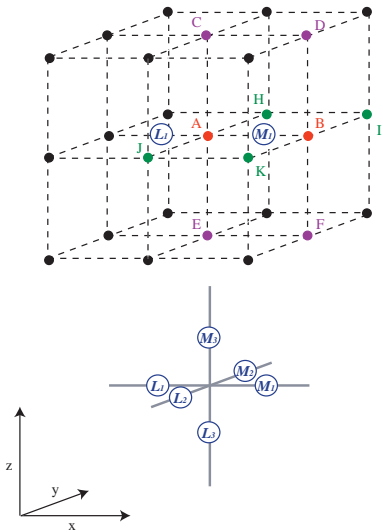}
	\caption{Discretisation scheme of the modified Poisson equation proposed in BM99. Density and potential is calculated on the grid nodes. The gradient components in $\mu(x)$ are estimated at the $L_i$ and $M_i$ points.}
	\label{fig:discr}
\end{figure}

Here is the discrete form of (\ref{eq:mondaqual}):
\begin{eqnarray}
\label{eq:monddiscret}
&\ &4\pi G \rho_{i,j,k}= \\ \nonumber
&\ &(\phi_{i+1,j,k}-\phi_{i,j,k})\mu_{M_1}-(\phi_{i,j,k}-\phi_{i-1,j,k})\mu_{L_1} \\ \nonumber
&\ &+(\phi_{i,j+1,k}-\phi_{i,j,k})\mu_{M_2}-(\phi_{i,j,k}-\phi_{i,j-1,k})\mu_{L_2} \\ \nonumber
&\ &+(\phi_{i,j,k+1}-\phi_{i,j,k})\mu_{M_3}-(\phi_{i,j,k}-\phi_{i,j,k-1})\mu_{L_3}) /h^2
\end{eqnarray}
with $\rho_{i,j,k}$ and $\phi_{i,j,k}$ the spatial density and potential discretized on a grid of step $h$, $\mu_{M_l}$, and $\mu_{L_l}$, the value of $\mu(x)$ at points $M_l$ and $L_l$ (Fig. \ref{fig:discr}).
The gradient component $(\partial /\partial x,\partial /\partial y,\partial /\partial z)$, in $\mu(x)$, are approximated by $({{\phi(B)-\phi(A)}\over h},{{\phi(I)+\phi(H)-\phi(K)-\phi(J)}\over {4h}},{{\phi(C)+\phi(D)-\phi(E)-\phi(F)}\over {4h}})$,

it is the stable numerical scheme proposed in BM99.

In the Newtonian case, the interpolation function is constant: $\mu(x)=1$, so that Eq. (\ref{eq:monddiscret}) becomes:
\begin{eqnarray}
4\pi G \rho_{i,j,k}&=&(\phi_{i+1,j,k}+\phi_{i-1,j,k}+\phi_{i,j+1,k}+\phi_{i,j-1,k}\\ \nonumber
&+&\phi_{i,j,k+1}+\phi_{i,j,k-1}-6\phi_{i,j,k})/h^2
\end{eqnarray}
We recognize the discrete Poisson equation with a 7-point Laplacian stencil.

To avoid the usual 2-body relaxation in simulated galaxies with insufficent number of particlees, the gravitational potential is softened through a convolution with a gaussian function ($\sigma=1.2$ cells). This value of the softening suppresses efficiently high spatial frequency noise, without introducing any bias (Dehnen 2001; Zhan 2006). For the following simulations, the calculation was made on a $257^3$ grid. The radius of the simulation box is $50\ kpc$, generally the galactic disc is truncated at $20\ kpc$. The same code can solve Poisson and Modified Poisson equations (it is just the coefficients of the PDE that are constant in Newton and variable in MOND).

\subsection{Boundary conditions}

In classical PM-code, fast Fourier transform (FFT) implies periodic boundary conditions. For isolated galaxy simulations, the interactions with periodic images can be suppressed using screening masses (James 1977). What kind of boundary conditions are possible to use for simulations of isolated galaxies with MG codes? This point is not trivial and is not developed in BM99 code. It is particularly important in MOND where the gravitational potential scales as $\log (r)$ far from the galaxy.
It might appear similar in Newton gravity with the dark matter halo, but the latter is nearly spherical in general, and the influence of mass exterior to the box is considered negligible. In MOND in the contrary, the potential at large distance is due to the baryonic disc (with spiral arms or bar).

Periodic conditions are not realistic at this scale and using a box eight time larger to suppress the periodic images is too costly in CPU time.
The most natural way is to use isolated boundary conditions. But this supposes we know the potential at the border of the box. To solve this problem we have to make an approximation. We use the MOND formula (Eq.(\ref{eq:deepM2N}) in the deep MOND regime), which links the MOND acceleration to the Newtonian acceleration. If $\mu(x)=x/(1+x)$, one has more generally:
\begin{equation}
a_\textrm{MOND}=a_\textrm{Newt} \nu\left({a_\textrm{Newt}/ a_0}\right)
\end{equation}
with
\begin{equation}
% \nu(x)={1\over 2}\left( 1+\sqrt{1+{4\over x}}\right) 
\nu(x)=0.5\left( 1+\sqrt{1+{4/ x}}\right).
\end{equation}
It is critical to use this expression directly in the Newtonian code principally because it is not true if the system has no symmetry (planar, cylindrical, or spherical). It is a concern even for an isolated galaxy. During its evolution, spiral arms and bars are formed and destroyed. The particle configuration is then not symmetric. However, we are interested only in the outer parts of the galaxy. We just need to determine the MOND potential on the boundary of the simulation box, that is far from the galactic center and its gravitational instabilities.

\begin{figure}
 \centering
 \includegraphics[width=8.5 cm]{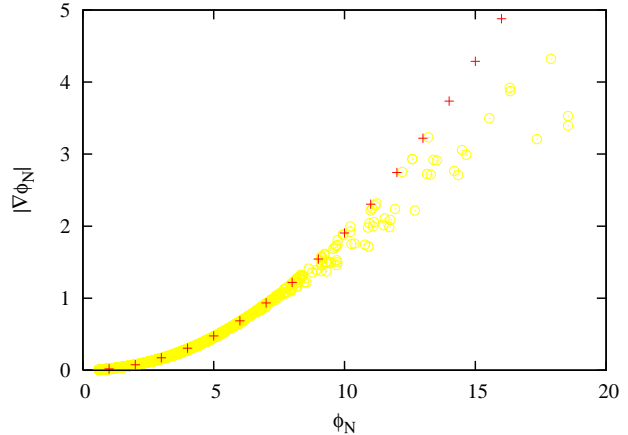}
 % test_approx.eps: 1048592x1048592 pixel, 300dpi, 8878.08x8878.08 cm, bb=
 \caption{Plot of $\mid\nabla\phi_{newt}\mid$ versus $\phi_{newt}$ for a barred galaxy with grand design spiral. The crosses represent \hbox{$\mid\nabla\phi_{newt}\mid$=$\phi_{newt}^2/GM$} (this is the spherical approximation where the galaxy is approximated by a point mass).}
 \label{fig:test_approx}
\end{figure}

Brada and Milgrom (1995) proposed a test to check if the approximation $a_\textrm{MOND}=f(a_\textrm{newt})$ is justified. They showed that $\mid\nabla\phi_{newt}\mid$ must be a function of $\phi_{newt}$ out of the disc. So by plotting $\mid\nabla\phi_{newt}\mid$ versus $\phi_{newt}$ at different positions in the box simulation we obtain Fig. \ref{fig:test_approx}. The approximation is good for low $\phi$ that is far from the galactic center. This is expected since the potential is more spherical. Hence, our solution to solve the boundary conditions problem is to compute the Newtonian potential by the FFT technique on a larger grid, then the Newtonian acceleration on each edge of the simulation box. Finally we use the MOND formula to deduce the MOND acceleration and compute the MOND potential on the border.

In this way, we obtain boundary conditions that are not fixed in time and that are not required to be homogeneous. We can take a small perturbation to the spherical symmetry like the disc or bar shape of the galaxy into account. Even if the correction is not very important to dynamical evolution, this makes the code more realistic.

\subsection{Tests}

We have made several tests to check the validity of the solution obtained by the MG technique: the analytical solution of a mass point, the Kuzmin disc. We present here a more demanding test for a totally non-symmetric system. It is the potential of a galaxy where a bar is formed during the simulation. We compute the MOND potential on the one hand with the MG technique and on the other hand with a classical relaxation scheme (NAG library). The second method is very inefficient (several hours when it takes less one minute with MG).
We plot the potential along the bar and perpendicular to the bar. The two different methods are in complete agreement. 
This plot also demonstrates the high symmetry at the outer boundary of the galaxy. Even if there is a bar ($5 kpc$), the potential at $25\ kpc$ is quasi-spherical (within a few $\%$).

\begin{figure}
 \centering
 \includegraphics[width=8.5cm]{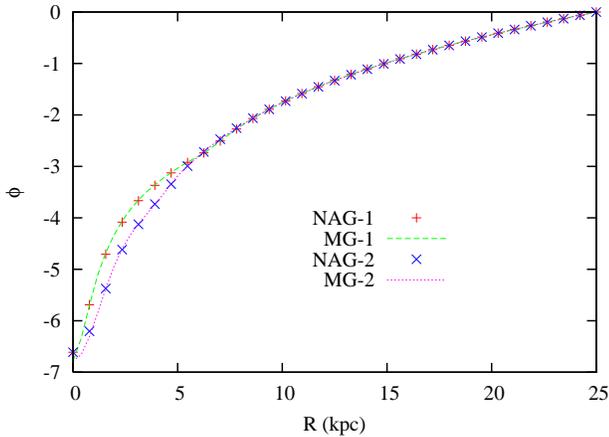}
 % test.eps: 1048592x1048592 pixel, 300dpi, 8878.08x8878.08 cm, bb=
 \caption{Test of the MG solver compared to the NAG routine \textit{D0ECF} to solve modified Poisson equation for a barred system with a grand design spiral structure; (1) potential perpendicular to the bar, (2) potential parallel to the bar.}
 \label{fig:testNAG}
\end{figure}

We have tested the dynamical evolution of a stellar disc in Newtonian gravity with the MG code compared to a classical FFT-PM-code. We obtained the same result for the time evolution of the bar. The code has been parallelized in open-MP since the G-S relaxation with red and black ordering allowed this. All red cells can be updated independently and this the same for the black cells.

\section{Analysis techniques}

\subsection{Fourier analysis in the galactic plane} 
The potential in the galactic plane is developed with the basis of the cosine function ($\Phi_m(r)$) and a phase term ($\phi_m(r)$):
$$
\Phi(r,\theta)=\Phi_0(r)+\sum_{m=1}\Phi_m(r)\cos[m\theta-\phi_m(r)]
$$
to calculate the maximum strength ($Q_m$) of the $m$ mode. We use the maximum force ratio:
$$
Q_m=max\bigg|{{F_{\theta,m}}\over{F_r}} \bigg|
$$
with the radial force:
$$
F_r=-{{d\Phi_0}\over{dr}}
$$
and the tangential force:

\begin{eqnarray*}
 max(F_{\theta,m}) &=& max\left( {{1}\over{r}}{{\partial\Phi}\over{\partial\theta}}\right)  \\
&=&{{m}\over{r}}\Phi_m(r).
\end{eqnarray*}

The bar strength is the maximum strength of the mode $m=2$. In general, the phase term, $\phi_2$, gives the rotation speed of the bar $\Omega_b$ and the derivative:
$$
\Omega_b={{\partial \phi_2}\over{\partial t}}
$$
But the mode $m=2$ could correspond to a two-arm spiral structure. Then, it is more informative to calculate the Fourier transform $\widehat{\phi_2} (r,\Omega)$ from $\phi_2(r,t)$. One can distinguish the angular velocity ($\Omega$) of a structure like a bar or a spiral arm versus the radius ($r$).
A bar is identified by a solid rotation in the central part of the galaxy:
$$\Omega(r)=cst=\Omega_b$$.

\subsection{Resonance}

We estimate the position of resonant orbits using the epicyclic approximation (Fig. \ref{fig:reso_n_m}). To do that, we need to determine the angular velocity of the stellar disc ($\Omega$),
$$
\Omega^2={{1}\over{r}}{{d\Phi_0}\over{dr}}
$$
the epicyclic frequency ($\kappa$),
$$
\kappa^2=r{{d\Omega^2}\over{dr}}+4\Omega^2
$$
and the vertical frequency ($\nu_z$),
$$
\nu_z^2={{\partial^2\Phi}\over{\partial z^2}}\bigg|_{z=0}.
$$

\subsection{Heating}

The heating of disc is computed by averaging the radial velocity dispersion, $\sigma_r(t)$, normalized by the initial $\sigma_{crit}$ (see Sect. \ref{sec:disp}) inside the $5\ kpc$ of the galaxy, giving the averaging Toomre parameter:
$$
Q_T=\langle {{\sigma_r}\over{\sigma_{crit}}} \rangle_{r<5\ kpc}
$$
with $\sigma_{crit}$, the critical velocity dispersion derived from the Toomre stability criterion,
$$
\sigma_{crit}={{3.36G\Sigma}\over{\kappa}}
$$
$\Sigma$ is the stellar surface density.

\subsection{Units}

In our code we use a unit system where the universal constant of gravity is: $G=1$, and the mass unity is $U_m=2.26\ 10^9\ M_\odot$. The length unity is $U_r=1.02\ kpc$ and the velocity unit is $U_v=100\ km.s^{-1}$. The time unit is $U_t=10\ Myr$. In this paper, when the units are not indicated, they are in this unit system.

\section{Initial conditions}

\begin{table}

\centering
\begin{tabular}{c c c c c c}
\hline\hline 
 Run & $M_d$ & $M_b$ & $a_d$ & $M_h$& $b_h$\\
\hline
Sa   & 40    & 12.65 & 4     & 206.4&14.8   \\
Sb   & 30    & 5     & 5     & 173.7&14.6   \\
Sc   & 20    & 2     & 6     & 148.8&14.5   \\
Sd   & 10    & -     & 6     & 129.7&13.1   \\
\hline
\end{tabular} 
\caption{Parameters. The characteristic length of the bulge for Sa, Sb, Sc galaxies is $1\ kpc$. For the characteristic height of the Miyamoto-Nagai disc, we choose $b_d/a_d=1/10$. The given mass is the truncated mass. The Toomre parameter value is the same in the DM and MOND model: $Q_T=2$. The disc is made of $2.10^5$ particles, the mass of the bulge particles is equal to the mass of the disc particle. The mass of dark matter particles is three times the mass of the disc particle.
}
\label{tab:param}
\end{table}

To study the stability differences for galaxies in the MOND and DM models, we construct a sequence of galaxies from early type to late type (Sa, Sb, Sc, Sd). Each type of galaxy corresponds to a set of two model galaxies, one for MOND and the other for DM. A galaxy, for a given type, has the same spatial density for the baryonic disc and bulge in the two models.
The stellar disc is modeled by a Miyamoto-Nagai disc:
$$
 \rho_d=\left( {{b_d^2M_d}\over{4\pi}}\right) {{a_dR^2+(a_d+3\sqrt{z^2+b_d^2})(a_d+\sqrt{z^2+b_d^2})^2}\over{\left[R^2+(a_d+\sqrt{z^2+b_d^2})^2 \right]^{5/2} (z^2+b_d^2)^{3/2}}} 
$$
with $M_d$ the mass of the disc (at infinity), $a_d$ and $b_d$ the characteristic length and height, and the bulge by a Plummer sphere:
$$
 \rho_b=\left( {{3M_b}\over{4\pi b_b^3}}\right) \left(1+{{r^2}\over{b_b^2}} \right)^{-5/2} 
$$
with $M_b$ the mass of the bulge (at infinity) and $b_b$ the characteristic length.

\subsection{Rotation curves}

From an observer's point of view, a galaxy must have the same rotation curve in MOND and in DM.
The shape of the rotation curve is imposed by the MOND model.
To obtain the same rotation curve (in the galactic plane) in DM we adjust a Plummer dark matter (live) spherical halo to fit the MOND rotation curve. Parameters of the dark matter halos are given in Table \ref{tab:param}. The error on the fit parameters is about $2\ \%$.

\subsection{Velocity dispersion}
\label{sec:disp}

We used the same value for the Toomre parameter ($Q_T$) for MOND and DM. The radial velocity dispersion is initialized by:
$$
\sigma_r=Q_T \sigma_{crit}.
$$
The tangential velocity dispersion ($\sigma_\theta$) is deduced from the epicyclic approximation,
$$
\sigma_\theta=\sigma_r{{\kappa}\over{2\Omega}}.
$$

\begin{figure}
 \centering
 \includegraphics[width=8.5 cm]{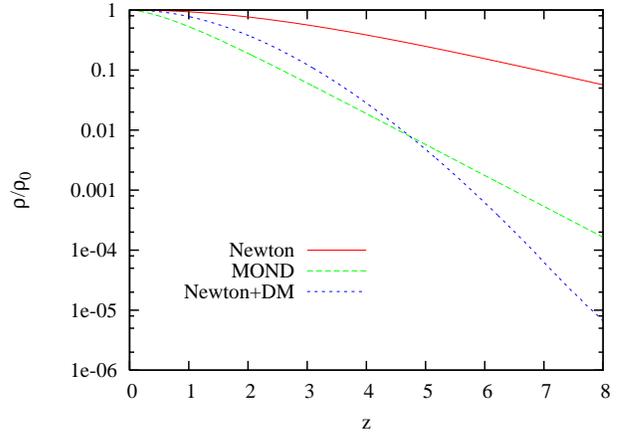}
 % p_vert.eps: 0x0 pixel, 300dpi, 0.00x0.00 cm, bb=50 50 266 201
 \caption{Vertical structure at the gravitational equilibrium of a disc in Newtonian gravity, MOND, and Newtonian with a dark matter halo.}
 \label{fig:p_vert}
\end{figure}

\noindent For the vertical velocity dispersion ($\sigma_z$), the hydrostatic equilibrium of an isothermal infinite stellar disc in Newtonian gravity gives:
$$
\sigma_z^2=H\pi G \Sigma(r)
$$
where $H$ is the characteristic height and $\Sigma(r)$ is the surface density.

We have calculated the vertical density profile ($\rho(z)$) in MOND for an isothermal infinite stellar disc (we consider the problem in one dimension). The equivalent pressure of the gas of star is $P=\rho \sigma_z^2$. The gravitational potential $\Phi$ is given by the modified Poisson equation. The gravitational equilibrium $(\nabla P= -\rho \nabla \Phi)$ is obtained when:

\begin{equation}
\label{eq:p_vert}
{{d}\over{dz}}\left[ \mu\left( {{\mid d\Phi/dz \mid}\over{a_0}} \right) {{d\Phi}\over{dz}} \right] =4\pi G\rho
\end{equation}
with
\begin{equation}
 {{d\Phi}\over{dz}}=-\sigma_z^2{{1}\over{\rho}}{{d\rho}\over{dz}}.
\end{equation}

$\sigma_z$ is a constant of $z$ but varies with $r$. 
We solved numerically Eq. \ref{eq:p_vert}, the result is plotted in Fig. \ref{fig:p_vert}. It shows also the vertical profile in Newtonian gravity: $\rho(z)=\rho_0\ \rm{sech}^2(z/H)$, and in Newtonian gravity with a dark matter halo. For this plot, we selected $\rho_0=2.3\ 10^{-6}\ M_\odot.kpc^{-1}$, which is a typical value of the outer disc. The dark matter halo is a Plummer sphere with $M_h=6\ 10^{11} M_\odot$ and $b_h=15\ kpc$ ($r=8\ kpc$). In our model we choose $H=cst$ in the DM model as well as in the MOND model. We want to keep the same initial height for a galaxy in DM an MOND. Figure \ref{fig:p_vert} shows that the vertical profile in MOND or in Newton with a dark matter halo are quite similar. The initial vertical velocity dispersion in MOND is the same as in the DM model. The stellar rotational velocity is not exactly the circular velocity ($v_c$), but $v_c-v_a$ where $v_a$ is the asymmetric drift deduced from Jeans equations applied to an infinitely thin disc. The system is relaxed initially in its axisymmetric potential to have a well stable virialized initial state.

\section{Results}
\subsection{Bar growth}
\subsubsection{Dark matter model}

\begin{figure*}
\centering
\includegraphics[width=18 cm]{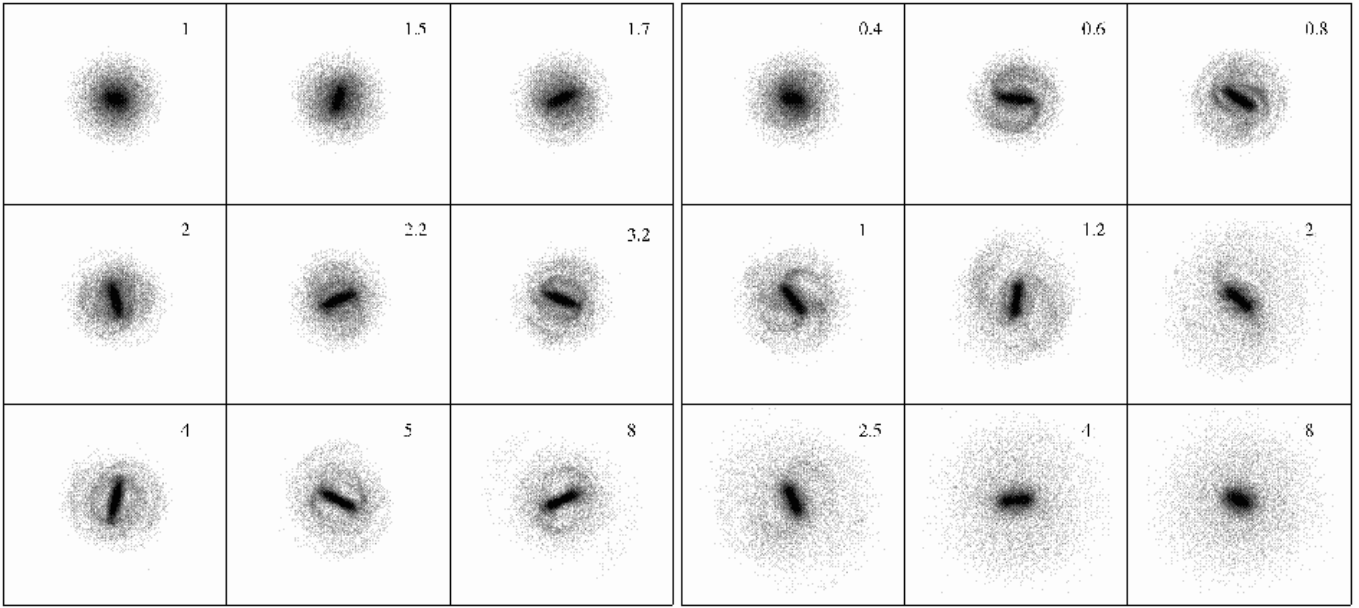}
\caption{Bar growth of Sa type ($Q_{T}=2$) in the DM model (left panel) and the MOND model (right panel). The size of the box is $80\ kpc\times 80\ kpc$. In the DM model, the bar develops in several Gyr. It can be noticed that the bar is surrounded by a ring at the end of the simulation. Particles are confined in the disc. In the MOND model, the bar appears quite rapidly (in less than $1\ Gyr$), and a lot of particles are spread out around it up to $30\ kpc$.}
\label{fig:snapshot_n_a}
\end{figure*}

\begin{figure*}
 \centering
 \includegraphics[width=8.5 cm]{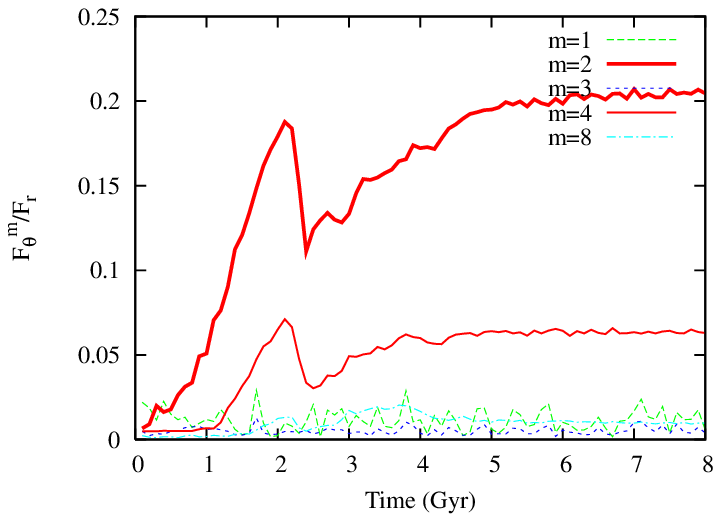}
 \includegraphics[width=8.5 cm]{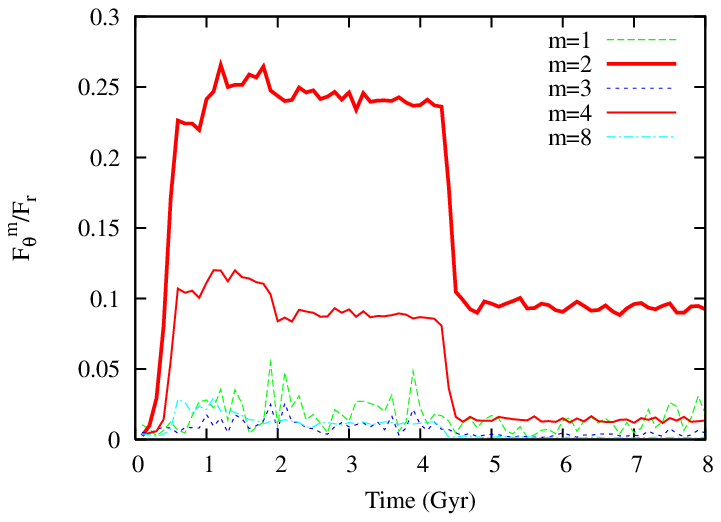}
 % mode_n_a.eps: 2500390x2500390 pixel, 300dpi, 21169.97x21169.97 cm, bb=
 \caption{Time evolution of $Q_m$ for $m=1,2,3,4,8$ of galaxy Sa in DM model (left) and in MOND model (right). In the DM model, the bar strength increases progressively compared to the MOND model where the bar reaches its maximum after $1\ Gyr$. The same drop appears at $t= 2.5\ Gyr$ in DM and $t=4.5 \ Gyr$ in MOND. After that, the bar strength increases again in the DM model, but not in MOND.}
 \label{fig:mode_n_a}
\end{figure*}

\begin{figure*}
 \centering
 \parbox{8.5 cm}{\includegraphics[width=8.5 cm]{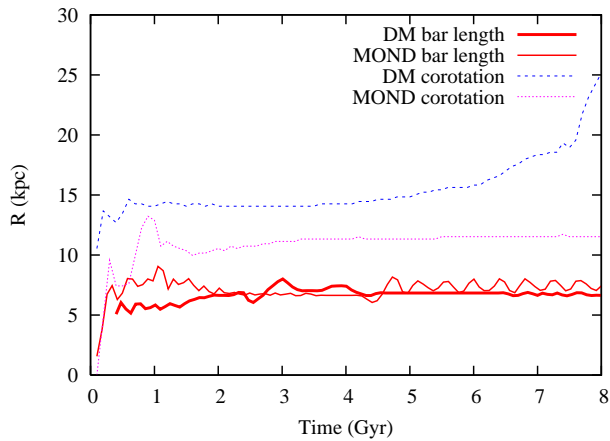}}\parbox{8.5 cm}{\caption{Bar length and corotation radius in the DM and MOND models. The bar length is defined by the radius where the bar strength ($Q_2(r)$) is equal to half its maximum. The difference between the end of the bar and the corotation radius in MOND is constant and smaller than in the DM model. The corotation radius in the DM model is shifted outward because the bar slows down (see Sect. \ref{sec:reso}).}\label{fig:bar_length}}
 
\end{figure*}

Figure \ref{fig:snapshot_n_a} (left panel) shows the evolution of an Sa galaxy in the DM model. At first sight, the initial Miyamoto-Nagai disc develops a bar instability in several $Gyr$. The bar length grows to $6\ kpc$ until $2\ Gyr$ (Fig. \ref{fig:bar_length}); its shape is rather squarish. For this run, we do not clearly see a grand design spiral structure during the bar growth. They exist, but they are more visible for a colder disc ($Q_{T}=1.5$). After $t=2\ Gyr$, transient spirals are developed between $10\ kpc$ and $20\  kpc$, while a ring appears at the end of the bar. The bar length continues to grow until $4\  Gyr$, as the ring extends, too ($6-7\ kpc$). During the period between $5\ Gyr$ and $8\  Gyr$, the bar changes its morphology and takes a butterfly shape. Its length does not increase contrary to the ring that has extended to $10\ kpc$.
Spirals that developed at $t=2\ Gyr$ have driven particles to a pseudo ring at $15\ kpc$, for $t=4\ Gyr$, to $25\ kpc$, at $t=8\ Gyr$. The nature of these rings will be discussed in Sect. \ref{sec:coro_ilr}.

Figure \ref{fig:mode_n_a} (left) displays the maximum strength of the bar as a function of time. One can distinguish three parts on this plot. First, $Q_2$ begins to increase until $2\ Gyr$. Then, this growth stops suddenly, the bar strength drops in $500\ Myr$. After $2.5\ Gyr$ the bar strength grows again until $5\ Gyr$ and appears to be constant until $8\ Gyr$.

\subsubsection{MOND model}

In MOND, the same galaxy (with the same value for $Q_T$) shows quite different structures (Fig. \ref{fig:snapshot_n_a}, right). First, a multi-spiral pattern appears rapidly after $0.4\ Gyr$ to give place to a grand design two-arm spiral at $0.6\ Gyr$. This spiral structure persists until $3\ Gyr$. The spiral arms have spread out particles up to $30\ kpc$. After $4\ Gyr$, the bar begins to be rounder and weakened. No rings are clearly visible in the MOND simulations.

$Q_2(t)$ is plotted on Fig. \ref{fig:mode_n_a} (right), the bar develops very soon ($1\ Gyr$), compared to the DM model. The bar strength is constant until $4.5\ Gyr$ where a drop occurs suddenly (like in DM at $2.5\ Gyr$). Afterwards, the bar strength remains low until the end of the simulation. However, the bar length is still constant ($6\ kpc$) even if the bar strength is weakened (Fig. \ref{fig:bar_length}).
\subsection{Pattern speed and resonance}
\label{sec:reso}

\subsubsection{Pattern speed}

\begin{figure}
 \centering
 \includegraphics[width=9 cm]{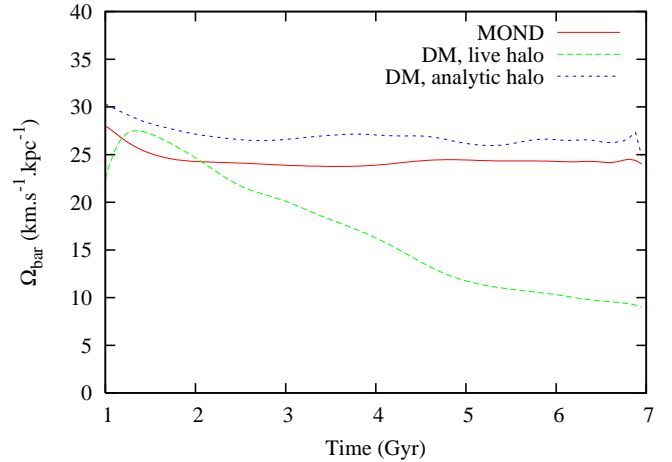}
 % vit_bar_Sa.eps: -1523272266x-417544648 pixel, 300dpi, -12897039.00x-3535211.25 cm, bb=
 \caption{Rotation speed of the bar for MOND and DM models. In MOND, the bar turns with the same velocity from the beginning to the end. In DM, particles of the halo slow down the rotation of the bar by dynamical friction.}
 \label{fig:vit_bar_Sa}
\end{figure}

The bar pattern speed is represented on Fig. \ref{fig:vit_bar_Sa}, still for the Sa-type galaxy ($Q_T=2$), for the MOND model, and for the DM model with a live and analytic halo.
The bar in DM with a live halo is considerably slowed down during the simulation ($25\ km.s^{-1}.kpc^{-1}$ to $10\ km.s^{-1}.kpc^{-1}$), while in MOND, the pattern speed is constant ($25\ km.s^{-1}.kpc^{-1}$). This plot emphasizes the dynamical friction effects experienced by the stellar bar against the DM particles. To confirm this result we perform a second simulation with the DM model using an analytical dark matter halo instead of a live halo. In this case, the pattern speed of the bar is still constant and corresponds to the MOND result.

\subsubsection{Corotation \& ILR}
\label{sec:coro_ilr}
\label{sec:units}

The bar pattern speed determines the position of resonant orbits in the reference frame of the bar rotating at $\Omega_b$. Because of velocity dispersion, stars do not just have a circular motion around the galactic center, they oscillate with the epicyclic frequency $\kappa$ (parallel to the galactic plane); likewise, they oscillate in $z$ with the frequency $\nu_z$. In most numerical simulations (with gas and dark matter) and in galaxies where it was possible to determine the bar pattern speed, the bar extends to its corotation (e.g., Buta \& Combes 1996).

\begin{figure}
 \centering
 \includegraphics[width=9 cm]{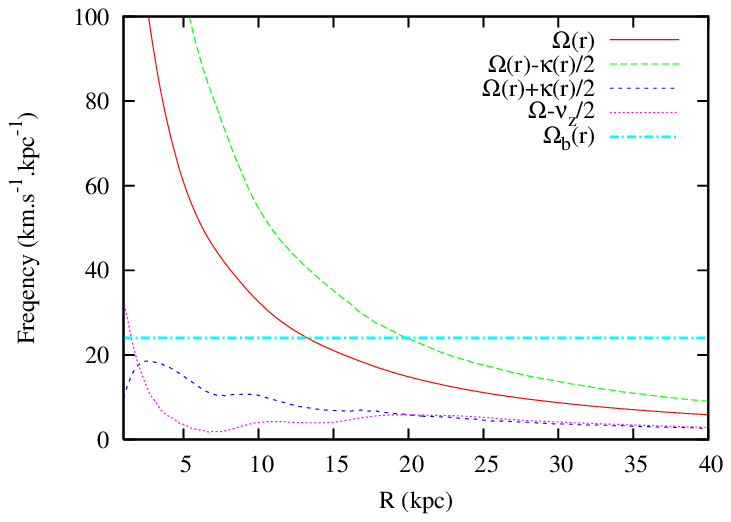}
 \includegraphics[width=9 cm]{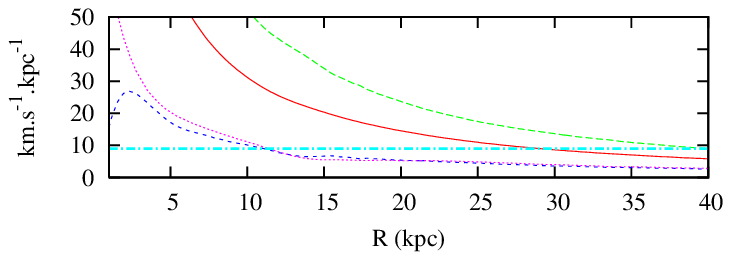}
 \includegraphics[width=9 cm]{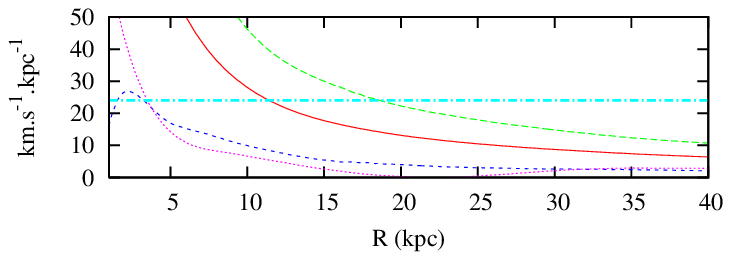}
 % reso_n_a_maxbar.eps: 1048592x1048592 pixel, 300dpi, 8878.08x8878.08 cm, bb=
 \caption{$\Omega(r)$ curve, and curves combined with the epicyclic frequency $\Omega_b=\Omega\pm\kappa/2$ and vertical frequency $\Omega_b=\Omega-\nu_z/2$: $t=2.5\ Gyr$ (top) and $t=8\ Gyr$ (middle) in the DM model, $t=8\ Gyr$ (bottom) in the MOND model.}
 \label{fig:reso_n_m}
\end{figure}

In the DM model, during the two first $Gyr$, the bar pattern speed is about $25\ km.s^{-1}.kpc^{-1}$ so that the corotation is nearly $13\ kpc$ (Fig. \ref{fig:reso_n_m}, top), while the bar ends at $7\ kpc$, but as the bar slows down, the corotation is shifted out to $20-25\ kpc$ at $t=8\ Gyr$ (Fig. \ref{fig:reso_n_m}, middle). Hence the ring surrounding the bar from $t=2\ Gyr$ until the end of the simulation does not correspond to the corotation resonance as it might be expected. The epicyclic approximation (Fig. \ref{fig:reso_n_m}, middle) indicates that we should have an outer and an inner Lindblad resonance (OILR, IILR), that is where $\Omega_b$ (the bar pattern speed) intercepts $\Omega\pm\kappa/2$.
The IILR is located very near the center of the galaxy. The OILR is about $12\ kpc$, where the ring is observed at the end of the simulation.
The bar appears to end nearly at the OILR.

\begin{figure}
 \centering
 \includegraphics[width=9 cm]{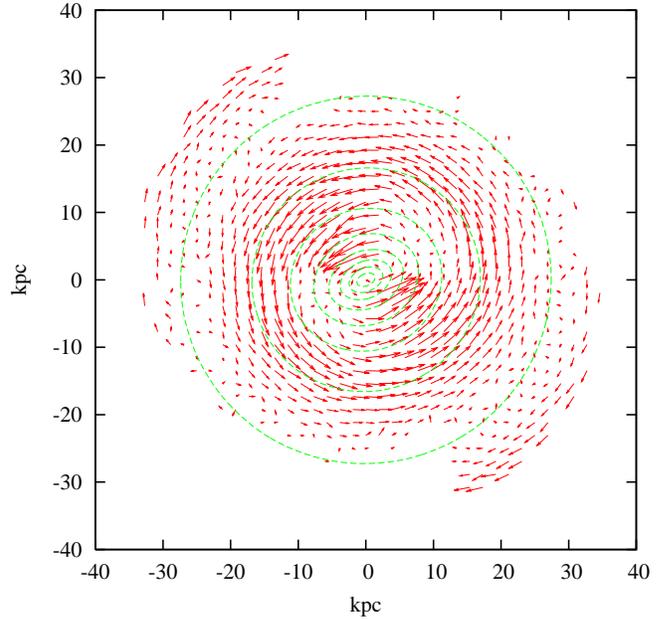}
 % vec.eps: 1048592x1048592 pixel, 300dpi, 8878.08x8878.08 cm, bb=
 \caption{Velocity field in the reference frame rotating with the bar. Vectors are parallel to the isopotential lines, indicating that the majority of orbits are $x_1$ type. }
 \label{fig:vect}
\end{figure}

Between the two ILR $x_2$, orbits must exist and destroy the bar. Figure \ref{fig:vect} displays the velocity field in the reference frame rotating with the bar, and the potential outline indicating the bar orientation. Corotation is well identified when vectors change orientation ($r\sim 25\ kpc$). The trajectory of particles are parallel to the bar potential like $x_1$ orbits, not perpendicular (like $x_2$ orbits). We have performed an orbit analysis to determine the existence of $x_1$ and $x_2$ orbits in this bar potential. The result is that the bar is not dominated yet by $x_2$ orbits. We have launched particles in the bar potential rotating at $\Omega_b=10\ km.s^{-1}.kpc^{-1}$. The value of the Jacobi's integral,
$$
E_J={{1}\over{2}}\dot{\mathbf{r}}^2+\Phi-{{1}\over{2}}\mid\mathbf{\Omega_b}\times\mathbf{r}\mid^2
$$
of particles varies between $h_{min}=-38$ (in our system units, $G=1$ see Sect. \ref{sec:units}), the bottom of the potential well, to $h_{max}=-13$, the potential nearly the corotation. $x_2$ orbits exist between $r=3-3.5\ kpc$ in an energy range about $-20<E_J<-18$. For lower $\Omega_b$ like $5\ km.s^{-1}.kpc^{-1}$, $x_2$ orbits appears clearly and more frequently between $r=2.5-9\ kpc$ in an energy range of $-26<E_J<-13$. Figure \ref{fig:reso_n_m} gives just an indication on the resonance with the epicyclic approximation. It can be noted that the drop in bar strength between $2\ Gyr$ and $2.5\ Gyr$ is correlated with ILR formation and its analogue in the z-direction (peanut formation, see the next section). In the MOND model the bar always ends near its corotation (Fig. \ref{fig:bar_length}), while the DM bar is relatively shorter.

\subsubsection{Vertical resonance and warp}

It can be shown that $\nu_z$ and $\kappa$ have a similar evolution. An equivalent resonance of the ILR exists in the z-direction if $\Omega_b=\Omega - \nu_z/2$. When particles resonate both in the plane and perpendicular to it, their vertical oscillations can be amplified and a peanut shape results.

\begin{figure*}[!]
 \centering
 \includegraphics[width=8.5 cm]{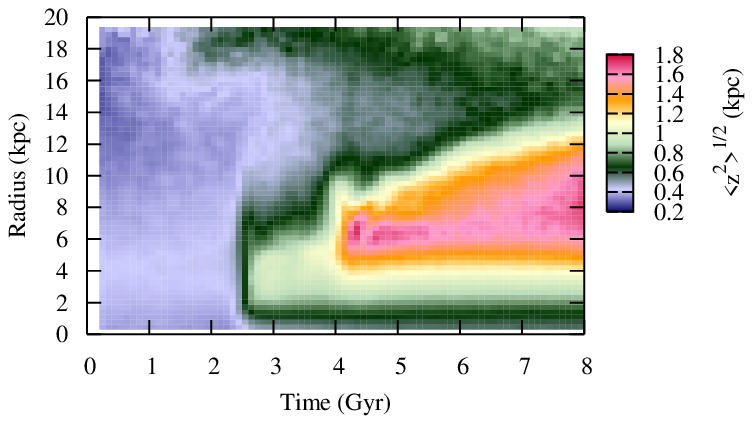}
 % z2_m_a.eps: 1048592x1048592 pixel, 300dpi, 8878.08x8878.08 cm, bb=
\includegraphics[width=8.5 cm]{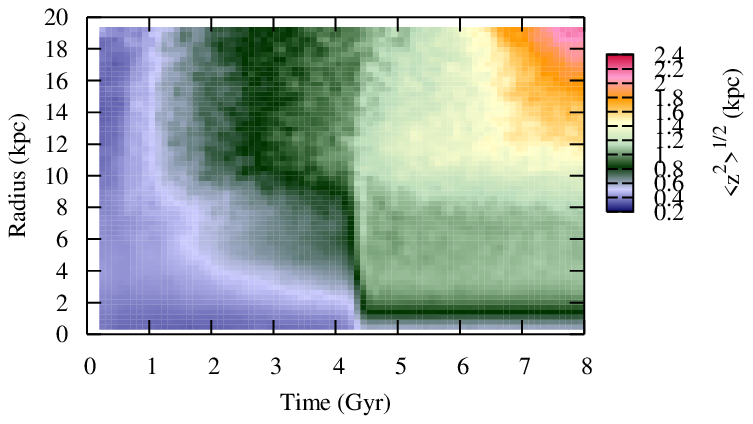}
 % z2_n_a.eps: 1095062x-1523265927 pixel, 300dpi, 9271.53x-12896985.00 cm, bb=

 \caption{Peanut formation in the DM (top) and MOND (bottom) models. These plots display $\langle z^2 \rangle^{1/2}$ of particles at several radii in a function of time. The peanut appears at $t=2.5$ in the DM model and $t=4.5$ in the MOND model. At these times, particles get out of the galactic plane in the range between $2-8\ kpc$. As the bar slows down in the dark matter halo, more and more particles resonate, and the peanut lobes extend to $12\ kpc$ at $t=8\ Gyr$. In the MOND model the peanut stays at the same place all along the simulation.}
 \label{fig:z2}
\end{figure*}

The drop in the evolution of the bar strength coincides with the peanut resonance, in DM as well as in MOND. At this moment particles are elevated out of the galactic plane. Stars are less bound and orbits become more oval, the bar strength is thus weakened.
Figure \ref{fig:z2} illustrates the moment when the peanut is formed. At $t=2.5\ Gyr$ in the DM model, and $t=4.5$ in the MOND model, particles between $2\ kpc$ and $8\ kpc$ resonate and are detached from the galactic plane.

\begin{figure}[!]
 \centering
\includegraphics[width=8.5 cm]{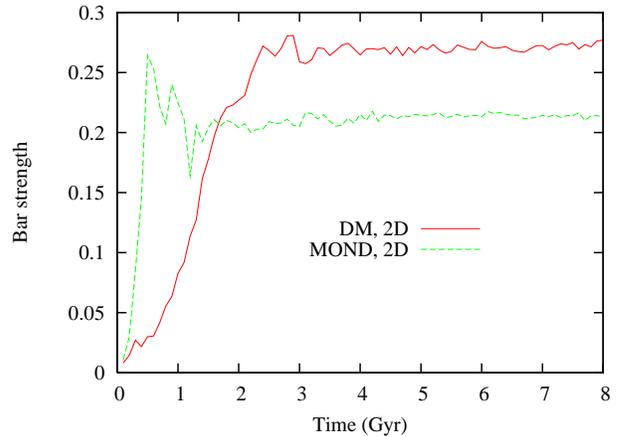}
 % bar_2D.eps: 1048592x1048592 pixel, 300dpi, 8878.08x8878.08 cm, bb=
 \caption{2D simulation of the Sa galaxy in DM and MOND models. The bar strength increases until a maximum and remains roughly constant.}
 \label{fig:bar_2D}
\end{figure}

\begin{figure}[!]
\begin{tabular}{cc}	
\includegraphics[width=4 cm]{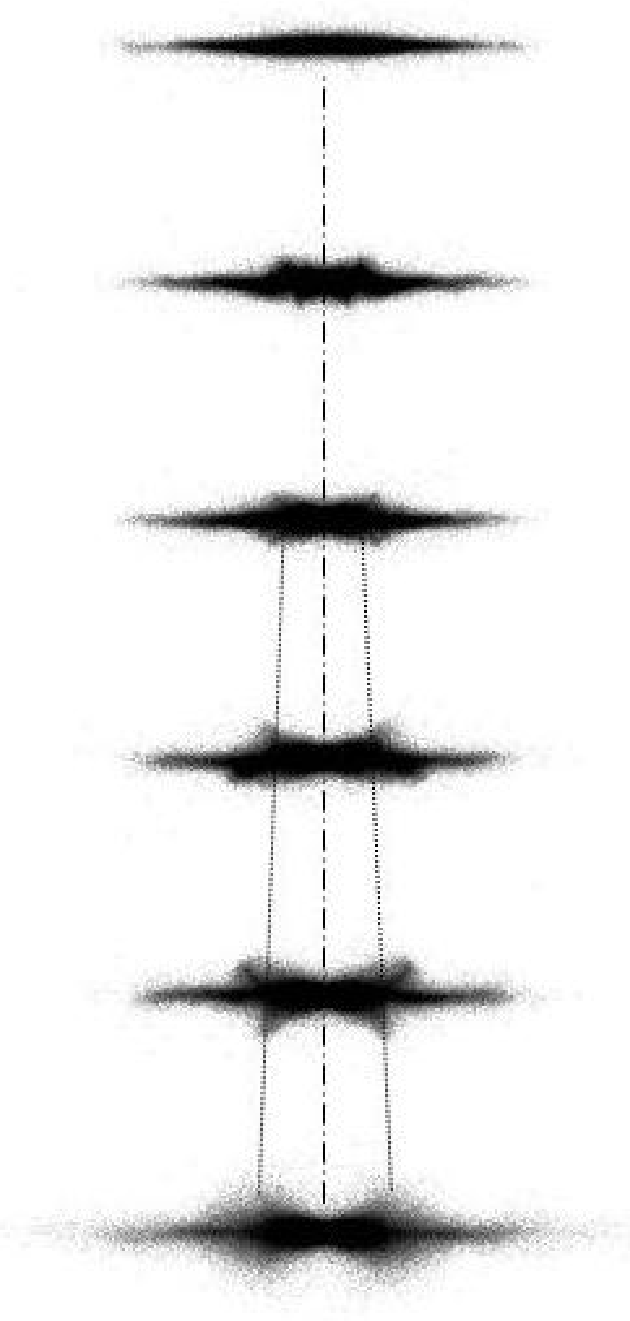}
\includegraphics[width=4 cm]{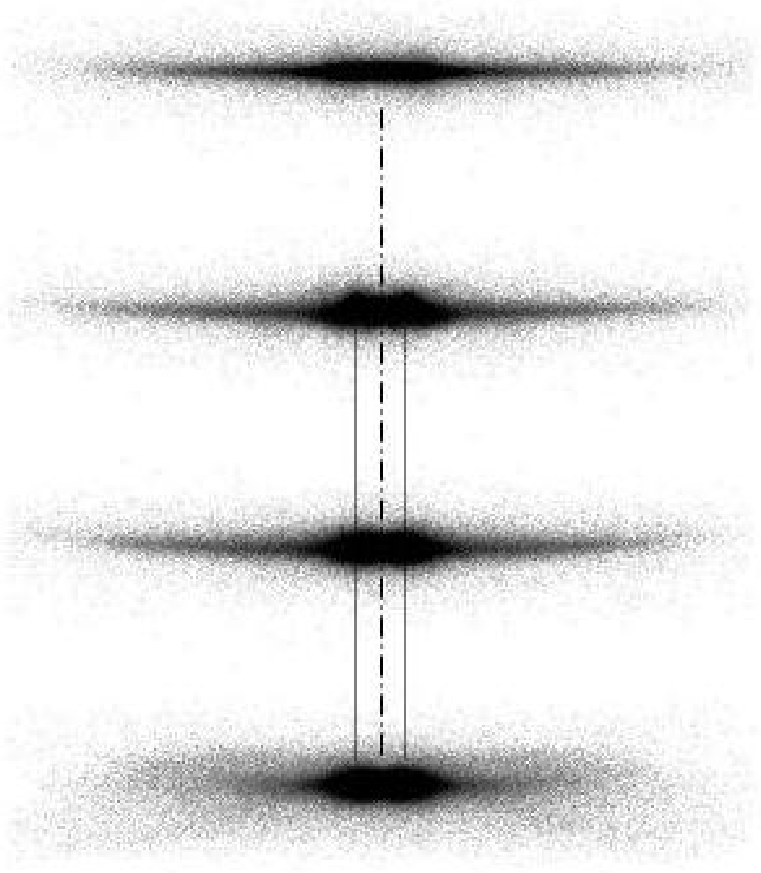}
\end{tabular}
\caption{Edge-on view of an Sa galaxy, shows the characteristic peanut shape. In the DM model (left), the position of the peanut lobes are radially shifted out as the bar slows down. In the MOND model (right), the peanut keeps the same size, one can notice the warp and flaring of the disc.}
\label{fig:n_a_p}
\end{figure}

To confirm that the drop in the bar strength is really due to the peanuts, we have performed a 2D-simulation for the DM and MOND models. In this case the bar strength is not weakened during its evolution Fig. \ref{fig:bar_2D}. There is no drop at $t=2.5\ Gyr$ in the DM model or at $t=4.5\ Gyr$ in the MOND model.

Like ILR and corotation in the DM model, the position of the z-resonance is also shifted out when the bar pattern speed decreases. In our simulation, dynamical friction acts quite progressively on the bar. The radial shifting of the peanut lobes is continuous. Martinez-Valpuesta et al. (2006) have observed similar phenomena in the formation of a peanut galaxy, but they distinguish two episodes.
They obtained a short bar due to chaotic orbits that appear between the ILR and vertical-ILR.

\begin{figure}
 \centering
 \includegraphics[width=8.5 cm]{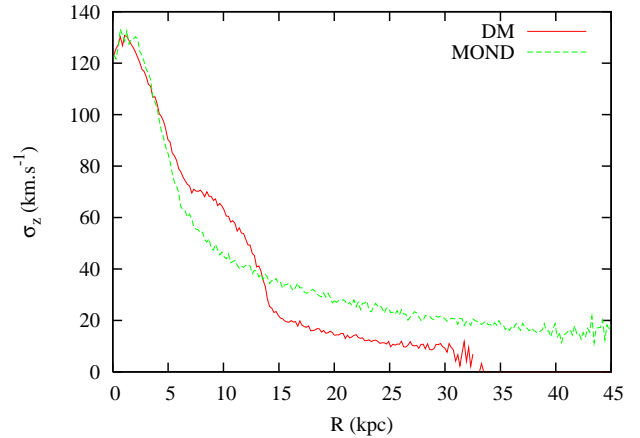}
 % sz.eps: -1235691336x-1235691336 pixel, 300dpi, -10462187.00x-10462187.00 cm, bb=
 \caption{Vertical velocity dispersion ($\sigma_z$) of the disc for the DM and MOND models at $t=8\ Gyr$. In DM, the peanuts makes $\sigma_z$ increase around $12\ kpc$. After $15\ kpc$, $\sigma_z$ in MOND is larger than in DM.}
 \label{fig:sz}
\end{figure}

Figure \ref{fig:n_a_p} shows several edge-on views of the galaxy in the DM and MOND models. In MOND simulation, the galactic disc is more easily warped than in DM. The disc begins to take a U-shape, to finally flare. The ratio $h/h_r$, where $h$ is the equivalent characteristic height and $h$ the characteristic length for an exponential disc, is about $0.26$ in the MOND model and $0.22$ in the DM model at $r=25\ kpc$. It is not as different as can be expected because $h_r$ in MOND is larger ($7\ kpc$) than in the DM model ($5.5\ kpc$). Particles are ejected radially further than in the DM model because of the angular momentum transfer (see Sect. \ref{sec:angmom}); the disc is thus less compact.
The origin of the flare comes from the vertical velocity dispersion that is more important in MOND for outer regions than in the DM model (Fig. \ref{fig:sz}). In MOND, vertical instabilities are developed because of the self-gravity, and the disc heats more. 

\begin{figure}
 \centering
 \includegraphics[width=8.5 cm ]{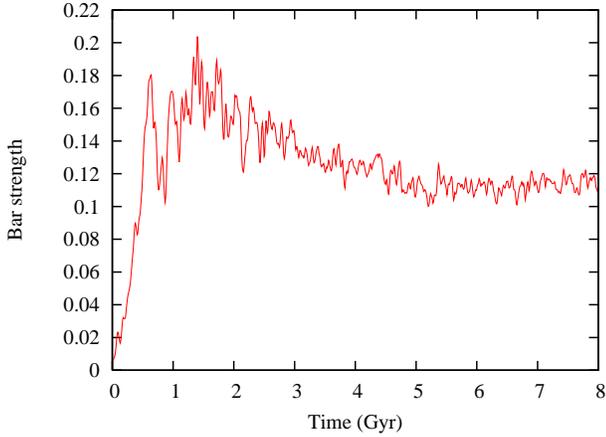}
 % bar_dest.eps: -19050x-1523265927 pixel, 300dpi, -161.29x-12896985.00 cm, bb=
 \caption{2D simulation of a cold stellar disc ($Q_T=1$) evolving in a small dark matter halo. The bar is formed quickly in a few galactic rotation. Bar weakens because of non-adiabatic bar growth (see text).}
\label{fig:bar_dest}
\end{figure}

We have seen that the weakening of the bar coincided with the peanut's occurrence. Let us note that the peanut is not the only way to weaken a bar in a pure stellar disc. If the disc is too cold, it develops a bar instability very quickly so that the stars have no time to follow a typical orbit with a constant bar strength. The corresponding disordered motions of the stars weaken the bar. This can be shown in a 2D (planar geometry) simulation (Fig. \ref{fig:bar_dest}).

\subsection{Heating}

\begin{figure}
 \centering
 \includegraphics[width=9 cm]{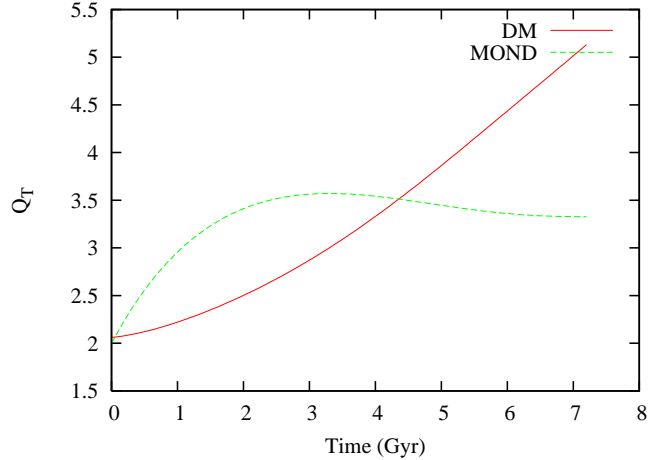}
 % heat.eps: 1048592x1048592 pixel, 300dpi, 8878.08x8878.08 cm, bb=
 \caption{Evolution of the Toomre coefficient value in function of time. The disc heats up in a few Gyr in the MOND model because of a complete self-gravity.}
 \label{fig:heat}
\end{figure}

The problem now is to understand why the bar strength continues to increase after the z-resonance in DM and not in MOND.
Part of the explanation can be found by following the evolution of $Q_T$ for the two models. The value of the Toomre coefficient indicates the heating rate of the disc. In these simulations, $Q_T$ starts at a value of $2$ in the whole disc. The evolution in the DM model and MOND model (Fig. \ref{fig:heat}) is differentiated from the beginning like the evolution of the bar strength.

\paragraph{Heating in the DM model.}

In the DM model, the disc heats progressively. When the peanut is forming, it weakens the bar and $Q_T\sim 2.7$. This value is not enough to avoid bar formation, which is why bar strength increases again. A disc in DM model needs a value of about $4$ for $Q_T$ to be stable and not form a bar.

\paragraph{Heating in the MOND model.}

In MOND, $Q_T$ increases to $3.5$ in a few Gyr. The apparition of a z resonance weakens the bar strength. At this time, the MOND disc is thus more stable because particles have more velocity dispersion. The bar strength does not increase anymore. We have performed another MOND simulation with $Q_T=3.5$ from the beginning. A weak bar is formed with a strength of about $0.12$. That corresponds with the bar obtained at the end of the simulation with $Q_T=2$ initially.

In MOND, all the matter participates to the dynamics; the galactic disc is completely self-gravitating, hence it heats up more than a disc rotating in a dark matter halo. 
The fact that no ring is clearly visible in the MOND simulation can be understood since the disc is hotter and might not sustain these features. The pitch angle of a density wave depends on $Q_T$. For a hot system, the theory predicts that the spiral will be more open than for a cold one; it is thus more difficult to form a ring.

\subsection{Angular momentum exchange}
\label{sec:angmom}
Another crucial point for the  bar formation is the exchange of angular momentum. For the bar to grow, particles of the disc have to lose angular momentum to fall in the inner region and have an elliptical orbit instead of a circular one. Angular momentum can be exchanged between the inner and outer parts. 

\paragraph{Angular momentum and dark matter.}

\begin{figure}
 \centering
 \includegraphics[width=8.5 cm]{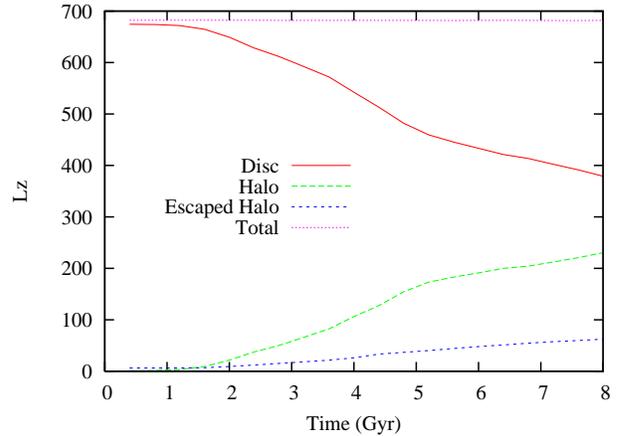}
 % Lz_n_a.eps: -14408668x-14934756 pixel, 300dpi, -121993.39x-126447.60 cm, bb=50 50 410 302
 \caption{Vertical component of the angular momentum in DM model. $L_z$ is exchanged from the disc to the halo during the bar formation. A non negligible proportion of particle (dark matter particles) escaped from the simulation box. Their motions are treated in a Keplerien potential as if all the matter inside the box is concentrated into a point mass.}
 \label{fig:Lz_n_a}
\end{figure}

\begin{figure}
 \centering
 \includegraphics[width=8.5 cm]{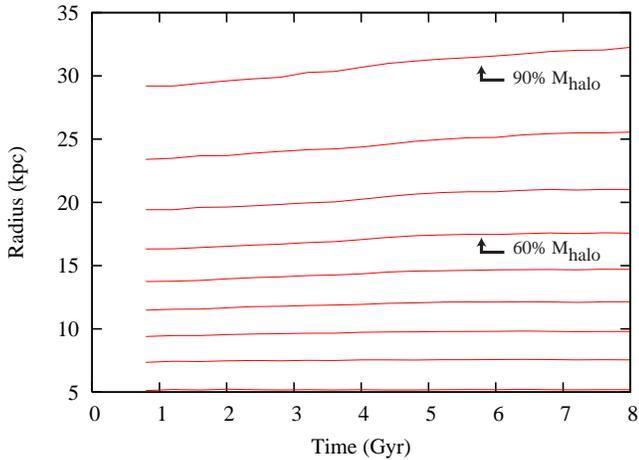}
 % masse_h_n_a.eps: 1048592x1048592 pixel, 300dpi, 8878.08x8878.08 cm, bb=
 \caption{In the DM model, the dark matter halo inflates because of angular momentum exchange from the disc to the halo.}
 \label{fig:masse_h_n_a}
\end{figure}

In DM, the halo can receive angular momentum from the disc. It is well illustrated in Fig. \ref{fig:Lz_n_a}: the disc loses about $30\%$ of angular momentum in the halo. In other terms if the halo increases its angular momentum, it will be less compact and will inflate. Figure \ref{fig:masse_h_n_a} represents the time evolution of several radii comprising a fraction of the mass between $10 \%$ to $90\%$ by $10\%$. One can notice the expansion of the radius below $60\%$. At $t=1\ Gyr$, $90\%$ of the mass is included in a sphere of $29\ kpc$; at $t=8\ Gyr$ this mass is in a sphere of $32\ kpc$ radius.

This exchange, which is efficient after $2\ Gyr$, contributes to bar formation (especially after the drop). Then the disc is not too hot and density waves can propagate. One can notice that the core of the halo is unaffected.

\paragraph{Angular momentum in MOND.}

\begin{figure}
 \centering
 \includegraphics[width=8.5 cm]{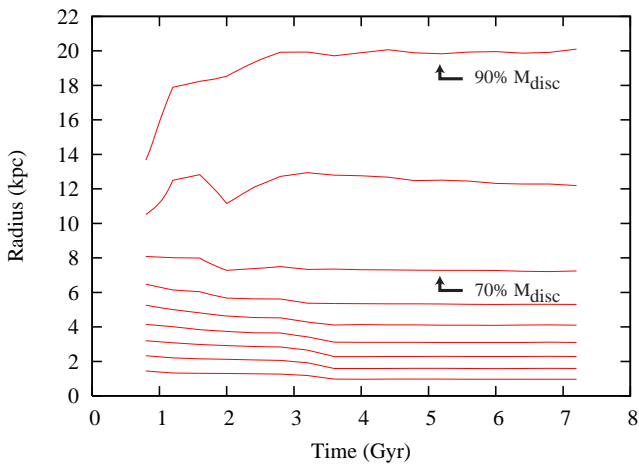}
 % masse_d_m_a.eps: 0x0 pixel, 300dpi, 0.00x0.00 cm, bb=
 \caption{In MOND, the outer part of the disc inflates while the center contracts. Angular momentum is transferred between these two regions during the bar growth.}
 \label{fig:masse_d_m_a}
\end{figure}

In MOND, the disc does not lose lot of angular momentum. Angular momentum is exchanged inside the disc itself. Figure \ref{fig:masse_d_m_a} shows the evolution of the same radius ($10-90\%$) of the mass versus time for the disc. The inner part of the disc loses angular momentum as expected because of bar formation (contraction of the disc below $8\ kpc$), and the outer part of the disc receives angular momentum from the inner region ($90\%$ of the mass is inside a $15\ kpc$ at the beginning and extends at $20 \ kpc$ at the end). This occurs during the first $3\ Gyr$, and the transfer is mediated by the spiral arms seen in Fig. \ref{fig:snapshot_n_a}. They evacuate angular momentum from the inner part to the outer part of the disc, and spread out particles around the disc. This is possible when the disc is not too hot. After this phase there is a saturation when the disc becomes stable and no density wave can propagate and increase the bar strength.

\subsection{Dark matter compared to MOND along the Hubble sequence}

\begin{figure*}
 \centering
 \includegraphics[width=8.5 cm]{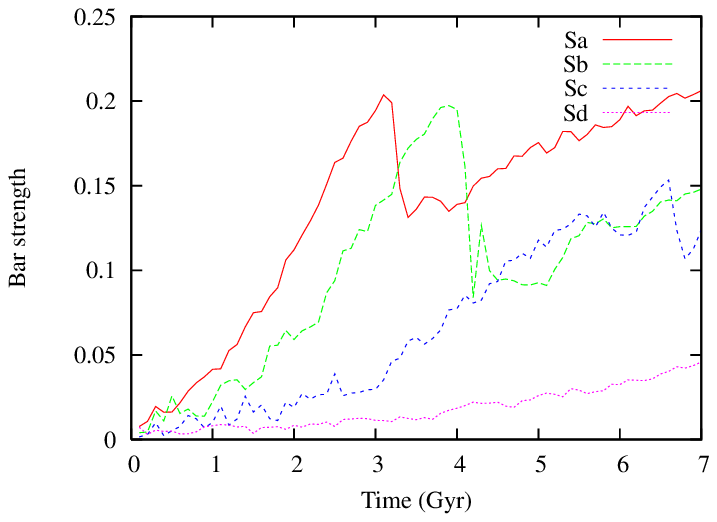}
 \includegraphics[width=8.5 cm]{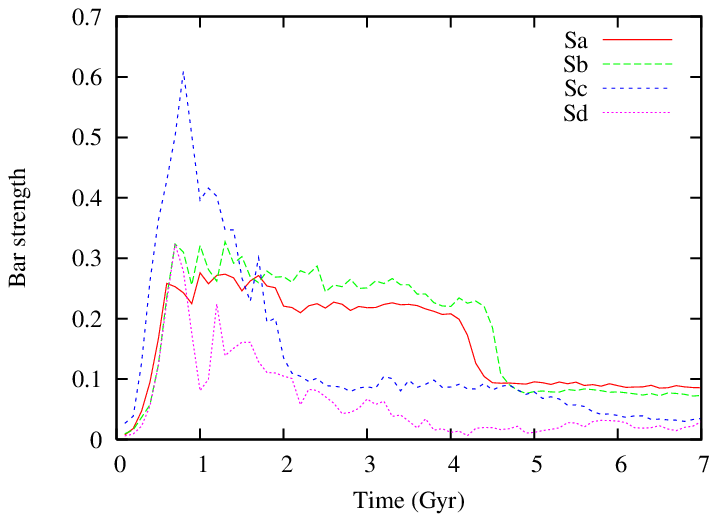}
 % mode_n.eps: 1048592x1048592 pixel, 300dpi, 8878.08x8878.08 cm, bb=
 \caption{Evolution of the bar strength in the DM (left) and MOND (right) models. Late-type galaxies in the DM model are more stable than early types; they need a larger proportion of dark matter to obtain the same rotation curve as in MOND. Late-type galaxies in MOND are too unstable; the bar destroys itself.}
 \label{fig:mode_n}
\end{figure*}

A series of simulations have been run to explore the parameters of galaxies along the Hubble sequence according to Table. \ref{tab:param}. In the DM model, from the early-type to late-type galaxies the ratio between the visible mass and the dark matter inside the optical radius increases. Thus, late-type galaxies are less self-gravitating than early types, so they are more stable (Fig. \ref{fig:mode_n}). 

In MOND, galactic discs are cold and form a bar in a few $Gyr$ whatever their type. The evolution scheme seen for the Sa type is reproduced for the Sb, Sc, and Sd type too. Even if the disc in MOND is cold and unstable at the beginning, it heats quickly and stabilizes itself along its evolution (Fig. \ref{fig:mode_n}).

Peanuts are formed for galaxies with a sufficiently massive bulge like Sa and Sb galaxies. In this case, the peanut occurrence weakens the bar. But in the DM model, the bar strength increases again because of the angular momentum transfer between the disc and the dark matter halo. While in MOND, the bar strength keeps low, and the disc heats up because of instability and stabilizes itself.

For Sc and Sd galaxies in the MOND model, the bar is formed too quickly (a few galactic rotations) because the system is too unstable. The stars have no time to settle in orbits supporting the bar at a given bar strength, since the orbital structure of the bar varies on a time scale shorter than the orbital period. These galaxies present a strong bar during a short time at the beginning of simulation to finish with a weak bar.

\begin{figure}
 \centering
 \includegraphics[width=8.5 cm]{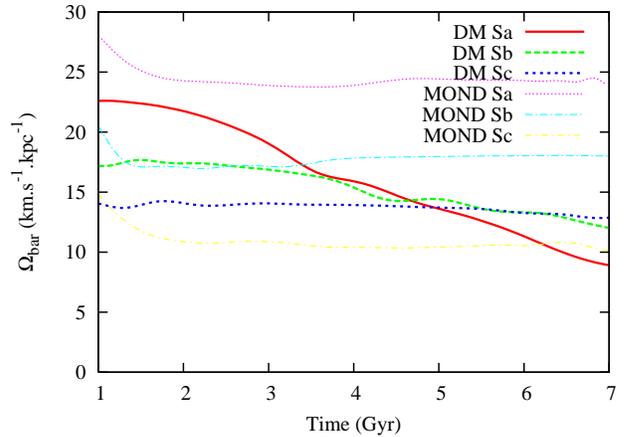}

 % vit_bar.eps: 1048576x0 pixel, 300dpi, 8877.94x0.00 cm, bb=
 \caption{Bar pattern speed in DM (left) and MOND (right) models. Dynamical friction effects in the DM model slow down the bar. In MOND pattern speed is still constant.}
 \label{fig:vit_bar_DM_MOND}
\end{figure}

The pattern speed of the bar is plotted in Fig. \ref{fig:vit_bar_DM_MOND}. In MOND the pattern speed is always constant for a given galaxy. Early-type galaxies have a higher bar pattern speed than late types (the disc is more massive). In DM, the bar is always slowed down by dynamical friction due to the halo. Late-type galaxies need more time to form a bar (Fig. \ref{fig:mode_n}), so their pattern speeds are less slowed down than for early types.

\begin{figure}
 \centering
 \includegraphics[width=8.5 cm]{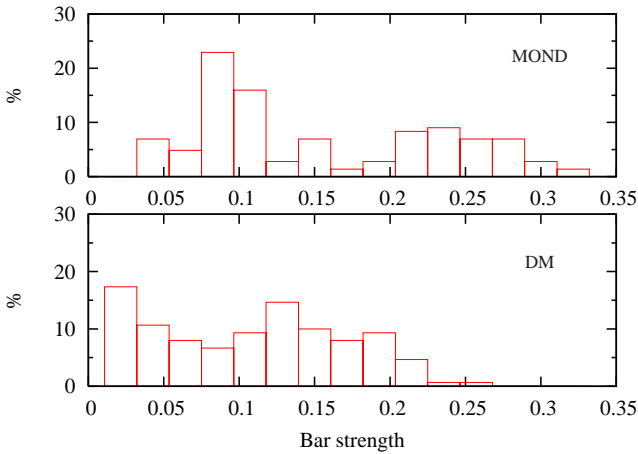}
 % bar.eps: 2500390x2500390 pixel, 300dpi, 21169.97x21169.97 cm, bb=
 \caption{Bar frequency in the simulated Hubble sequence in MOND (top) and in DM (bottom). Bars are stronger with MOND and there is a dearth of galaxies without bars in MOND, but not in the DM model. }
 \label{fig:freq_bar}
\end{figure}

We have made a statistical study of the bar strength for typical galaxies of the Hubble sequence. Figure \ref{fig:freq_bar} shows the bar frequency obtained using the time spent by a galaxy with a given bar strength. Two tendencies are clear from this plot. First, galaxies in the MOND model have stronger bars ($Q_2>0.25$) than in the DM model. The MOND discs are more unstable at the beginning so they form a strong bar very quickly. Secondly, there is a hole at low bar strength in the MOND model that is not present in the DM model. This is due to the dark matter halo that stabilizes the disc at the beginning, so it takes more time to a galaxy to form a bar. The bar strength distribution obtained from the observations presents some characteristics that are reproduced with the MOND model. In particular, there is a small proportion of galaxies with a very weak bar, and a few galaxies have very strong bar (e.g., Block et al. 2002 and Whyte et al. 2002).

\section{Discussion and conclusion}

In this paper, the dynamical evolution of pure stellar discs in MOND is compared to Newtonian gravity with DM, using numerical simulations. We have developed an N-body code that solves the modified Poisson equation in three dimensions using MG technique for the potential solver. The simulations in the DM models have been performed with the same code by solving Poisson equation with the same MG technique. 

For isolated galaxy evolution, the main difference between the MOND gravity and the Newtonian gravity with dark matter is the self-gravity of the disc. Even if the acceleration in MOND scales as $M^{1/2}$ instead of $M$ in Newtonian gravity (BM99), the dark matter halo in the DM model stabilizes more efficiently the disc. From a given initial state, the MOND disc is more unstable than the DM disc in the sense that it develops a bar instability sooner, for the same Toomre parameter value.

One of the main effects of the dark matter halo is the dynamical friction experienced by the stellar bar against the DM particles. The bar pattern speed is slowed down in the DM model. This does not exist in MOND. The bar pattern speed in MOND keeps constant all along the evolution, thus higher than in the DM model. This has consequences on the position of the resonances like corotation. Bar lengths are often compared to the corotation radius. In this case, bars obtained with MOND end closer to the corotation radius.

The 3D simulations reveal several differences between MOND and Newtonian gravity with dark matter. Peanuts are formed in the DM model as well as in the MOND model, but peanut lobe positions, which correspond to the z-inner Lindblad resonance, depend on the bar pattern speed. In MOND, the peanut always remains the same size ($\Omega_b=cst$), contrary to the DM model where the lobes are radially shifted far from the center (about $12\ kpc$). In MOND, successive instabilities due to self-gravity make the vertical velocity dispersion higher, in the outer region of galaxies, than in the DM model. There is a higher tendency for MOND discs to warp and flare.

Two mechanisms to weaken a bar have been described. First, for galaxies with a massive bulge (early type) a peanut resonance can be formed. This vertical motion of stars dilutes the bar concentration in the plane and makes the bar strength decrease. Secondly, if the disc is cold and unstable, it forms a bar so quickly that the orbital structure of the bar varies on a time scale shorter than the orbital period, and the stars cannot settle on orbit supporting the bar.

The present simulations reveal that the dark matter halo has two contradictory influences on the disc stability. On the one hand, the DM halo stabilizes the disc and delays the bar formation; on the other hand, it can reinforce the bar growth when the bar is forming by accepting the angular momentum from the disc stars, in particular after the peanut's formation. In contrast, peanut galaxies in MOND should have low bar strength. 

Statistically, the MOND bar frequency corresponds better to the observations than to the DM model. Indeed, there is a hole in the barred galaxy distribution for low bar strength and more galaxies distributed at high bar strength. But in this work, only stellar discs are considered without any gas component. Bar formation and destruction is affected by the gas component in the spiral galaxies. In particular gas accretion allows galaxies to have several bar cycles (Bournaud \& Combes, 2002).  Gas components will be added in future works. 

Through this work, we help to develop numerical tools for testing MOND. Using this code, many physical situations could be simulated. More complex systems will be studied, such as interacting galaxies where MOND might reveal larger differences compared with the DM model.

%%%%%%%%%%%%%%%%%%%%%%%% acknowledgments
\begin{acknowledgements}
We are grateful to B. Semelin and F. Bournaud for helpful discussions. Simulations in this work have been carried out with the IBM-SP4 of the CNRS computing center, at IDRIS (Palaiseau, France)
\end{acknowledgements}
%%%%%%%%%%%%%%%%%%%%%%%%%%%%%%%%%%%%%

%%%%%%%%%%%%%% references %%%%%%%%%%%%%%%%
{}
%%%%%%%%%%%%%%%%%%%%%%%%%%%%%%

\end{document}